# Extraordinary epitaxial alignment of graphene islands on Au(111)

Joseph M. Wofford[1,2], Elena Starodub[3], Andrew L. Walter[4,5], Shu Nie[3], Aaron Bostwick[4], Norman C. Bartelt[3], Konrad Thürmer[3], Eli Rotenberg[4], Kevin F. McCarty[3], Oscar D. Dubon[1,2]

[1] Department of Materials Science and Engineering, University of California, Berkeley, California 94720, United States
[2] Materials Sciences Division, Lawrence Berkeley National Laboratory, Berkeley, California 94720, United States
[3] Sandia National Laboratories, Livermore, California 94550, United States
[4] Advanced Light Source, Lawrence Berkeley National Laboratory, Berkeley, California 94720, United States
[5] Department of Molecular Physics, Fritz-Haber-Institut der Max-Planck-Gesellschaft, Faradayweg 4-6, 14195 Berlin, Germany

**Abstract**

Pristine, single-crystalline graphene displays a unique collection of remarkable electronic properties that arise from its two-dimensional, honeycomb structure. Using *in-situ* low-energy electron microscopy, we show that when deposited on the (111) surface of Au carbon forms such a structure. The resulting monolayer, epitaxial film is formed by the coalescence of dendritic graphene islands that nucleate at a high density. Over 95% of these islands can be identically aligned with respect to each other and to the Au substrate. Remarkably, the dominant island orientation is not the better lattice-matched 30º rotated orientation but instead one in which the graphene [01] and Au [011] in-plane directions are parallel. The epitaxial graphene film is only weakly coupled to the Au surface, which maintains its reconstruction under the slightly *p*-type doped graphene. The linear electronic dispersion characteristic of free-standing graphene is retained regardless of orientation. That a weakly interacting, non-lattice matched substrate is able to lock graphene into a particular orientation is surprising. This ability, however, makes Au(111) a promising substrate for the growth of single crystalline graphene films.

Advances in the epitaxial growth of graphene have transformed this remarkable material from a testbed for two-dimensional, solid-state physics [1, 2] into a technological material in advanced electronics [3] and sensing [4]. Rapid developments in understanding the nucleation and growth behavior of graphene, particularly on SiC [5, 6] and various metal substrates [7, 8], have bridged the gap between fundamental studies and applications. However, further investigations regarding epitaxial growth focused specifically on elucidating the interplay between the substrate and graphene structure, both physical and electronic, remain critically important. To this end, we have investigated the formation of graphene on Au(111) from vapor-deposited elemental carbon using *in-situ* low-energy electron microscopy and diffraction. On this surface graphene islands grow with homogeneous orientations and coalesce into a highly ordered epitaxial film that displays the characteristic linear electronic dispersion associated with free-standing graphene.

To date, the growth of graphene on copper foils by chemical vapor deposition (CVD) has garnered significant attention as it yields large-area, monolayer films [9-11]. These films consist of domains, or grains, as large as several hundred micrometers across, each distinguishable by the particular rotation of its crystalline lattice — originating from the spread of relative crystallographic orientations adopted by individual islands as they nucleate and grow on the Cu foil [12, 13]. In this context Au represents a unique opportunity for a comparative study of substrate materials. Like Cu, Au has a low C solubility, an important substrate attribute for growing monolayer films [14], and does not form binary phases with C. It has a much lower vapor pressure than Cu leading to less motion of substrate steps during growth, which has been shown to complicate graphene growth [12, 15]. Oxide formation - which can be deleterious - is mitigated by the relatively chemically inert Au, while the resulting inefficiency of hydrocarbon decomposition at the Au surface is readily circumvented via growth using an elemental C source. From the standpoint of practical growth procedures for applications, Au (111)-textured foils can be fabricated with grains centimeters in size [16] and electrochemical methods can be used for the separation of graphene from metal surfaces [17] thereby preserving the foil for subsequent re-use as a substrate.

Graphene was grown on Au(111) single crystals under ultra-high vacuum conditions. This allowed for the real-time monitoring and analysis of the sample during growth by both surface-sensitive low-energy electron microscopy (LEEM) and low-energy electron diffraction (LEED). Carbon was deposited from an electron-beam-heated graphite rod at substrate temperatures ranging from 770 ºC to 940 ºC. Prior to C deposition, the Au surface was cleaned and smoothed using O and Ar sputtering in conjunction with annealing to temperatures up to 1000 ºC. *Ex-situ* X-ray photoemission spectroscopy (XPS) of the graphene-Au surface was performed at the Advanced Light Source (ALS), Lawrence Berkeley National Laboratory. Spectra reveal no contaminants with only C and Au present.

Graphene islands appear nearly instantaneously upon exposure of the Au surface to a C flux, nucleating homogeneously on Au terraces as well as along Au step edges as is shown in Figure 1a. (For a movie of graphene island growth, see Supplementary Video 1.) The average overall nucleation density of ~$7.5 \times 10^8$ cm$^{-2}$ is substantially higher than that reported on other metal surfaces [7, 9, 12, 18, 19]; for instance, on Cu(001), four-lobed islands nucleate at a density of ~$1.3 \times 10^6$ cm$^{-2}$, more than two orders of magnitude lower than that observed on Au(111) [12]. Islands on Au terraces form single-domain, dendritic shapes with ramified branches of approximately equal length. In contrast, those nucleated on step edges expand more rapidly along the Au step, resulting in an elongated form. This accelerated growth often results in ribbons of continuous graphene along Au step bunches prior to the formation of a full film. Figure 1a contains examples of both types of islands. The dendritic morphology of graphene islands on Au(111) is similar to that observed on Cu surfaces [12, 15, 20].

With continued C deposition the discrete graphene islands coalesce to form a complete film, as shown in Figure 1b. Even upon the formation of a continuous film, no evidence for a second graphene layer was found. Current-voltage (I-V) LEEM and LEED measurements both confirmed this, while X-ray photoemission spectroscopy (not shown) from a partial film precluded the possibility of other C phases (such as amorphous or carbidic C) being present, which might not generate sufficient diffraction contrast to be easily visible in LEEM. Within the conditions examined, the graphene-Au(111) heterostructure is stable at just one monolayer, resisting the formation of further layers or the accumulation of additional C on the sample surface.

LEED analysis shows the graphene to be of high quality, with sharp, six-fold symmetric diffraction patterns. Diffraction also reveals that there is a single dominant in-plane graphene orientation relative to the Au(111), single-crystal substrate. Moreover, in this orientation the lattices of the graphene and Au(111) surface are aligned. (The graphene[01] and Au[011] in-plane directions are parallel.) We label this orientation R0, as illustrated in Figure 1d. Small fractions of the graphene film are rotated by 30º relative to the Au lattice (R30, Figure 1e) or randomly oriented. The high degree of orientational homogeneity in these films contrasts with graphene grown on Au(111) by a modified CVD process [21], which displays extensive rotational disorder (revealed by LEED) not only in the relative fraction of R0- and R30-oriented domains but also in the variation of alignment within each of these orientations; the 2 sets of diffraction arcs generated by the distribution of relative orientations in the CVD case are comparable to those resulting from the scatter around graphene's two degenerate configurations on the (001) surface of Cu [12]. The exceptionally clean substrate surfaces, combined with the relatively slow growth rates used here likely contribute to this increased ordering. From the perspective of classical thin-film epitaxy, which favors the formation of commensurate films, the prevalence of the R0 orientation is puzzling – a graphene lattice in this orientation must be strained by 17% strain to be commensurate with the Au(111) lattice. In contrast, R30 graphene has only a relatively modest mismatch of ~1.5%. Indeed, the R30 orientation has been emphasized in theoretical studies [22, 23] because of its smaller mismatch and has been proposed to be the more stable epitaxial orientation [24]. However, we argue below that the interaction between graphene and Au is so weak that commensurability considerations are not applicable.

The dominance of the R0 orientation is striking, as demonstrated in the LEEM micrographs in Figure 2a-d. A sub-monolayer graphene coverage conveniently allows for contrast between the islands and exposed Au substrate, and the set of over 450 individual graphene islands in these particular images enables a thorough statistical sampling of the graphene-Au(111) structure. A comparison between the bright-field LEEM image (Figure 2a,c) and a dark-field image of the same region taken in the first order diffraction condition of R0 graphene (Figure 2b,d) shows that over 95% of graphene islands have the R0 orientation.

To further examine this materials system we have mapped the electronic structure of epitaxial graphene on Au(111) using angle-resolved photoemission spectroscopy (ARPES) at the electronic structure factory endstation (SES-R4000 analyzer) at beamline 7 of the ALS. For the ARPES spectra a photon energy of 95 eV was used, giving overall resolutions of ~25 meV and 0.01 Å$^{-1}$. The spatial area sampled by ARPES was typically 50 to 100 μm. Measurements were performed at room temperature, and the pressure was <2x10$^{-10}$ Torr. The resulting Fermi surface, as shown in Figure 3a, confirms the distribution of graphene orientations identified by LEED. The ARPES spectra in Figure 3 have low-background signals and narrow line widths. As the line widths of ARPES spectra are directly related to the defect scattering rate, this indicates a clean, well-ordered graphene layer. Additionally, the Fermi surface of the sample shows few states that are not attributable to the R0 or R30 graphene, further demonstrating the structural homogeneity of the sample. The characteristic linear electronic dispersion of graphene is unmistakable in the band structure plots along the Γ-K direction. The dispersion in Figure 2b is extremely close to that of free-standing, high-quality graphene with no evidence of defect induced deterioration (e.g., grain boundaries). Interestingly, the linear dispersion near the charge neutrality point is preserved for both the R0 and R30 orientations along their respective Γ-K directions, despite the substantially different local configuration of C atoms relative to the underlying Au (Figures 1d,e and 3b,c). The structure of the Fermi surface and the preservation of graphene's linear dispersion are both indicative of a graphene film of high crystalline quality that is decoupled from the Au substrate. This weak coupling between graphene and Au is consistent with theoretical calculations [24] as well as experimental investigations of Au-intercalated graphene [25-27]. That orientational disorder has been shown typically to be inversely proportional to the interaction strength between graphene and a metal surface [28] makes the dominance of the R0 alignment in this weakly interacting system even more surprising [29].

Graphene films supported by metal substrates generally display charge transfer with the metal surface, and both the R0 and R30 orientations on Au(111) are no exception (Table 1). Indeed, the Fermi level shift superimposed on the high-quality, defect-free ARPES spectrum of graphene on Au(111) is the most noteworthy evidence of the interaction between the two materials. This charge transfer with the Au surface results in a slight *p*-type doping with a hole concentration of ~$6.2 \times 10^{11}$ holes/cm$^2$, which is in good agreement with first principle predictions [24]. Only graphene on Ir(111) displays a lower level of charge transfer; however, in that case the electronic structure of graphene is complicated by interactions with the Ir bands close to the Fermi level. As these interactions are absent in graphene on Au(111), this system is well suited for investigations of quasi-freestanding graphene, an attribute that is being investigated in transferred and Au-intercalated graphene [25, 30-32].

The weak interaction between the graphene and Au surface also manifests itself in the nanoscale structure of the two materials, as revealed by scanning tunneling microscopy (STM). The dominant feature in the STM image of R0 graphene on Au(111) (Figure 4a) is a hexagonal array of depressions separated by about 1.8 nm. This array results from the moiré between the graphene and the topmost Au atoms. The STM image contains a more-subtle feature, faint ribbons that bend from nearly horizontal in the upper left to nearly vertical in the lower right. These ribbons are the distinct "herringbone" surface reconstruction adopted by bare Au(111) [33, 34]. Thus, the graphene-covered Au surface has the same distinctive reconstruction as clean Au(111).

The ordering of a bare Au surface into the herringbone structure as the temperature drops below 865K [35] is the result of a delicate energy balance [36] that, remarkably, is not tipped by the presence of a graphene overlayer. Thus, the presence of the herringbone reflects the exceptional degree to which the graphene and Au surface are decoupled. Under a small tunneling current (< 10 pA) both the Au herringbone and the moiré between the Au and the graphene can be observed simultaneously, revealing small distortions in the moiré as the graphene crosses the Au dislocations, which form the chevron-like herringbone. All of these features can be accounted for without invoking any lateral distortions in the graphene, as the following simple atomic model demonstrates (Figure 4b). For this model we generated the Au(111) surface employing a 2-dimensional Frenkel-Kontorova algorithm [37]. A unit cell of the Au herringbone reconstruction is shown in the upper part of Figure 4b. A graphene lattice is placed onto the Au surface without allowing the graphene sheet to relax laterally. To simulate the STM image we compute the brightness as a function of the lateral positions of the C atoms with respect to the underlying Au lattice (e.g., on-top positions are bright and 3-fold hollow sites are dark as described in Ref. [38]) as well as the amount of local distortion in the Au lattice. The simulation reveals the chevron-like ribbons that originate from the Au herringbone reconstruction and the hexagonal array of depressions from the graphene/Au moiré. In fact, the periodicities (1.8 nm) of the experimental and simulated moirés are the same within measurement error. The ability of this model to reproduce the main features of the STM image without laterally distorting the graphene sheet is further testimony to the minimal influence exerted on the graphene film by the Au substrate, and vise-versa. Raman spectroscopy provides further evidence that the graphene is essentially unstrained on Au(111) [39].

Comparison between these results and recent reports of graphene grown on Au reveals significant differences that are likely integral to the growth process used. Previous efforts to grow graphene on Au involved the use of molecular precursors as the C source, in one case ambient-pressure methane CVD [40] and, in the other, surface irradiation with ionized species derived from the electron bombardment of ethylene [21]. In the former case, depending on growth conditions, films vary from multi-layer, defective graphene to films with regions of variable thickness. In the latter case, incomplete monolayer films display high local structural quality in STM; however, as mentioned previously, LEED reveals extensive rotational disorder. The limited catalytic activity of the Au surface could serve to restrict the decomposition of molecular precursors and their incorporation into graphene. Thus, physical vapor deposition (PVD) of C provides a clear advantage over CVD-based processes for the growth of graphene on weakly catalytic surfaces such as Au. PVD is a scalable deposition technology and can be used to explore otherwise inaccessible growth conditions.

In conclusion, graphene films of sufficient orientational homogeneity to be quasi-single crystalline can be grown on Au(111). The dominant approach that until now has guided efforts to grow high-quality epitaxial graphene films centers on increasing graphene island size (i.e., lowering the nucleation density). The physical vapor growth of epitaxial graphene on Au(111) provides an alternative mechanism: film formation via the coalescence of small, highly aligned islands (i.e., increasing the nucleation density).

**Acknowledgments**

Work at the Advanced Light Source was supported by the Director, Office of Science, Office of Basic Energy Sciences, of the U.S. Department of Energy under Contract No. DE-AC02-05CH11231. Work at Sandia was supported by the Office of Basic Energy Sciences, Division of Materials and Engineering Sciences, U.S. Department of Energy under Contract No. DE-AC04-94AL85000. ODD acknowledges support from the National Science Foundation (Grant No. DMR-1105541). JMW acknowledges support from a National Science Foundation Graduate Research Fellowship. ALW acknowledges support from the Max Planck Society.

**Figures**

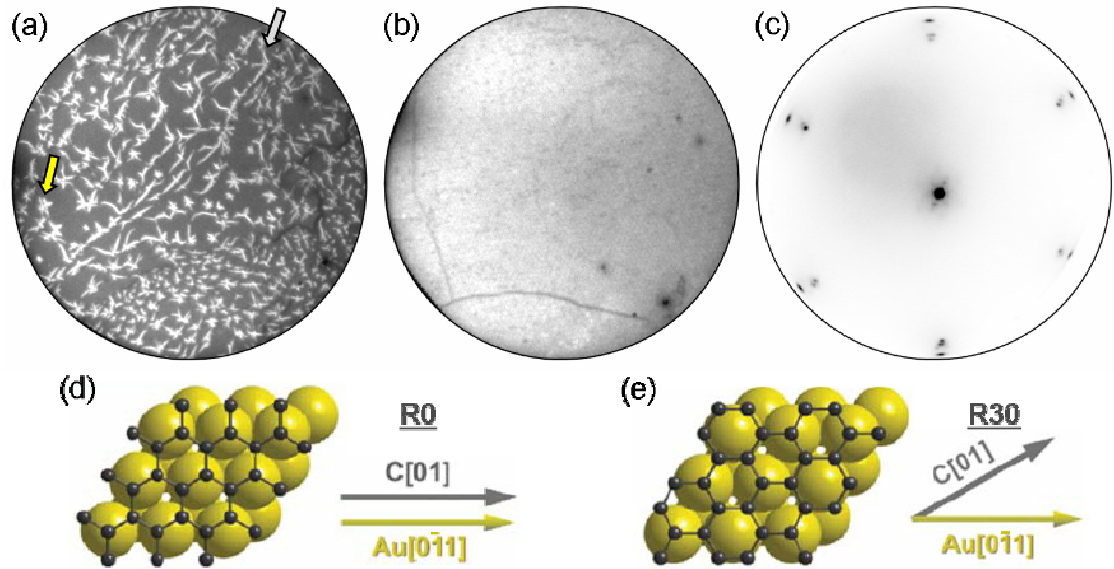

**Figure 1.** *In situ* **observation of graphene growth by LEEM.** (**a**) LEEM image of graphene islands (bright) on Au(111) (dark). Islands nucleate both on Au terraces (e.g., at the yellow arrow) and at the edges of Au steps (e.g., at the gray arrow) and develop dendritic shapes during growth at 880 °C. (**b**) LEEM micrograph after the islands have grown to completely cover the substrate. The broad, dark lines are likely wrinkles caused by differential thermal contraction between the graphene and Au during cooling from the growth temperature. (**c**) Selected area LEED showing that the graphene orients preferentially in an R0 alignment with a small minority of domains rotated by 30°. Diffraction from the Au(111) herringbone surface reconstruction is evident. (**d**) and (**e**) Atomic models of the R0 and R30 orientations, respectively, where a carbon atom is arbitrarily assumed to lie atop a Au atom. The field of view of the LEEM micrographs is 9 μm.

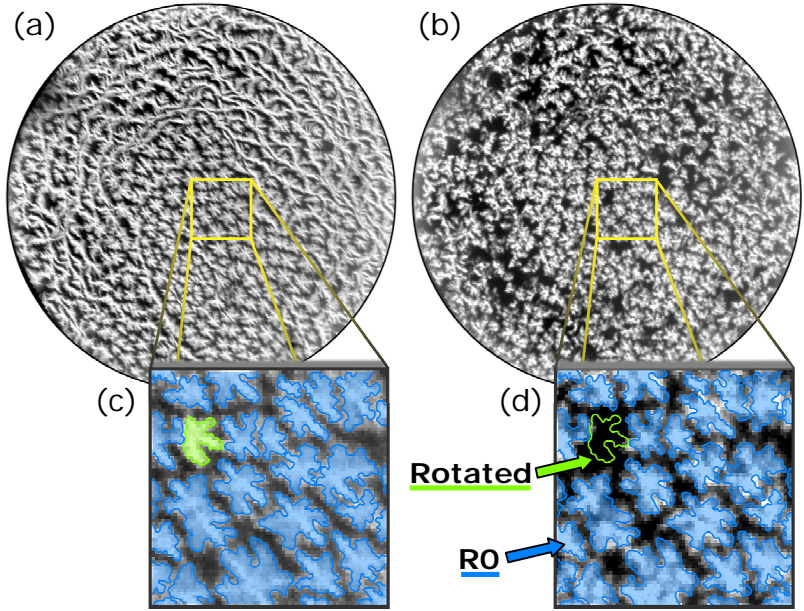

**Figure 2. Rotational homogeneity of graphene on Au(111) by dark-field LEEM.** (**a**) Bright-field LEEM micrograph of ~0.75 monolayer of graphene on Au(111) grown at 880 °C. (**b**) Dark-field LEEM micrograph of the same region formed from a first order diffraction spot of R0 oriented graphene. Islands with the R0 orientation are visible in both images, while those with any other orientation are illuminated in **a** but not in **b**. This distinction is demonstrated in the expanded bright-field (**c**) and dark-field regions (**d**) (field of view = 1.2 μm). In both images the R0 oriented islands, outlined and tinted blue, are bright while the single island of a different orientation, outlined and tinted green, is dark in **d**. Over 95% of islands in the 9 μm field of view of images **a** and **b** have the R0 orientation, resulting in a quasi-single crystal film upon their coalescence.

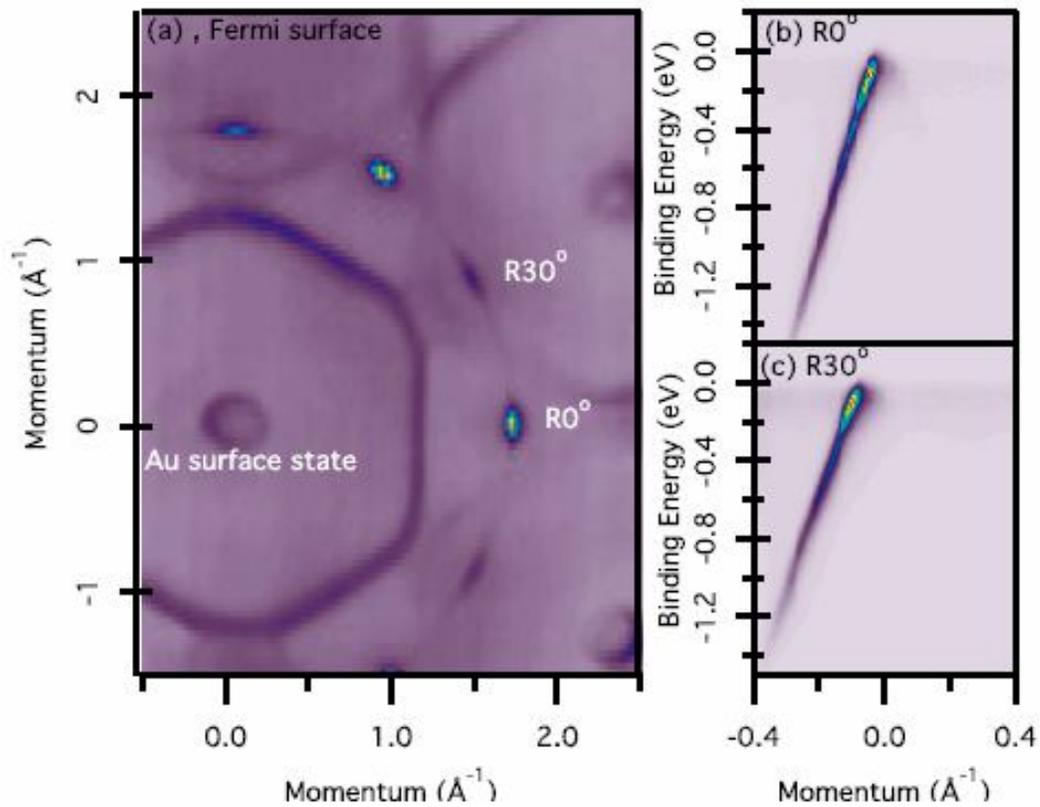

**Figure 3. Fermi surface and band-structure maps of graphene on Au(111) by ARPES.** The Fermi surface distribution in **a** shows the dominance of the R0 orientation, with a small fraction of graphene domains at R30. Charge transfer with the Au substrate depresses the graphene Fermi level, resulting in a slight *p*-type doping with a concentration of ~6.2 x $10^{11}$ holes/cm$^2$. Notice the persistence of the Au surface state, indicative of a clean Au surface. Spectra along Γ-K for the R0 (**b**) and R30 (**c**) orientations show that both maintain the characteristic linear dispersion of free-standing graphene.

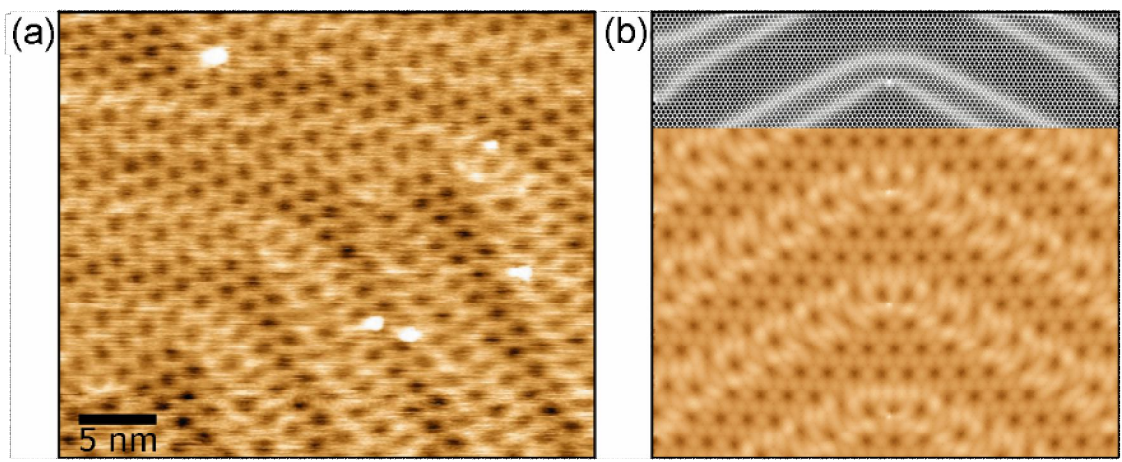

**Figure 4. Surface structure of graphene on Au(111) by STM.** (**a**) 35 nm x 29 nm STM image of R0 graphene. The hexagonal array of depressions, ~0.15 Å deep and ~1.8 nm far apart, is due to the moiré between the uppermost Au atoms and the graphene layer. Superimposed are faint ribbons 6 nm apart, which bend from nearly horizontal in the upper-left part of the image to nearly vertical in the lower-right part. The ribbons have the same dimensions as the "herringbone" dislocation pattern of the reconstructed Au(111) surface. (**b**) Simulation of the STM image (**a**) based on a simple atomic model. The Au(111) surface [as exposed region is shown at the top of **b**] is covered by a laterally undistorted R0-oriented graphene lattice. The brightness of the model surface is determined by the distortions of the Au lattice as well as the lateral position of the C atoms with respect to the Au substrate (e.g., on-top positions are bright). That graphene-covered Au develops the same dislocation structure as the bare Au(111) surface highlights how little the gold is affected by the presence of graphene.

| Substrate | $k_f$ (Å$^{-1}$) | $\rho$ (cm$^{-2}$) | doping type | $E_{gap}$ (eV) |
|---|---|---|---|---|
| F-SiC [41] | -0.119 | 4.5x10$^{13}$ | hole | |
| H-SiC [26] | -0.043 | 5.9 x10$^{12}$ | hole | |
| Au-SiC [26] | -0.015 | 7.2 x10$^{11}$ | hole | |
| **Au(111)** | **-0.014** | **6.2 x10$^{11}$** | **hole** | |
| Au-Ni(111) [25] | -0.014 | 6.2 x10$^{11}$ | hole | |
| Au-Ru(0001) [27] | -0.014 | 6.2 x10$^{11}$ | hole | |
| Ir(111) [42] | -0.013 | 5.4 x10$^{11}$ | hole | |
| Cu-Ni(111) [25] | 0.035 | 3.9 x10$^{12}$ | electron | 0.18 |
| Ag-Ni(111) [25] | 0.056 | 1.0 x10$^{13}$ | electron | 0.32 |
| Cu(111) [43] | 0.060 | 1.1 x10$^{13}$ | electron | 0.25 |
| 6√3 C-SiC [26] | 0.075 | 1.8 x10$^{13}$ | electron | 0.00 |
| Ru(0001) [27] | 0.095 | 2.9 x10$^{13}$ | electron | 0.00 |

**Table 1. Literature values for the Fermi vectors ($k_f$), doping level ($\rho$), doping type and gap size ($E_{gap}$) determined from ARPES measurements on various substrates.** The graphene on Au(111) considered here is highlighted in bold. The doping level is determined using the size of the Fermi surface via the relation $\rho=k_f^2/(2\pi)$; gaps are provided for the *n*-doped samples only. Substrates labeled X - Y have X as an interface layer between the graphene and the substrate Y. Graphene on Au has almost as low a doping level as on Ir but maintains its characteristic band structure, making it an ideal candidate for studies of quasi-freestanding graphene.